\documentclass[aps, prb, twocolumn, dvipsnames, showpacs, superscriptaddress,longbibliography]{revtex4-1}
\usepackage{graphicx, xcolor}
\usepackage{float}
\usepackage{amsmath} 
\usepackage{hyperref}
\usepackage{mathtools}
\hypersetup{
        colorlinks,
   citecolor=blue,
   filecolor=blue,
   linkcolor=blue,
   urlcolor=blue,
   breaklinks=true}

\usepackage[utf8]{inputenc}
\usepackage{braket}

\newcommand\ii{\mathrm{i}} 
\makeatletter
\newcases{mycases}{\quad}{%
  \hfil$\m@th\displaystyle{##}$}{$\m@th\displaystyle{##}$\hfil}{\lbrace}{.}
\makeatother

\newcommand\PRLSection[1]{{\textcolor{blue}{\textit{#1: }}}}  
\newtheorem{conj}{Conjecture}

\begin{document}
\title{Fate of the non-Hermitian skin effect in many-body fermionic systems}
\author{Faisal Alsallom}
\thanks{These two authors contributed equally}
\affiliation{Department of Physics, Massachusetts Institute of Technology, Cambridge, Massachusetts 02139, USA}
\affiliation{Institute of Physics, Ecole Polytechnique Fédérale de Lausanne (EPFL), CH-1015 Lausanne, Switzerland}

\author{Lo\"{i}c Herviou}
\thanks{These two authors contributed equally}
\affiliation{Institute of Physics, Ecole Polytechnique Fédérale de Lausanne (EPFL), CH-1015 Lausanne, Switzerland}

\author{Oleg V. Yazyev}
\affiliation{Institute of Physics, Ecole Polytechnique Fédérale de Lausanne (EPFL), CH-1015 Lausanne, Switzerland}

\author{Marta Brzezi\'{n}ska}
\email[Corresponding author: ]{marta.brzezinska@epfl.ch}
\affiliation{Institute of Physics, Ecole Polytechnique Fédérale de Lausanne (EPFL), CH-1015 Lausanne, Switzerland}

\date{\today}

\begin{abstract}
    We revisit the fate of the skin modes in many-body non-Hermitian fermionic systems. 
    Contrary to the single-particle case, the many-body ground state cannot exhibit an exponential localization of all eigenstates due to the Pauli exclusion principle.
    However, asymmetry can still exist in the density profile, which can be quantified using the imbalance between the two halves of the system.
    Using the non-Hermitian Su-Schrieffer-Heeger (SSH) chain as an illustration, we show the existence of two distinct scaling regimes for the imbalance.
    In the first one, the imbalance grows linearly with the system size, as generically expected.
    In the second one, the imbalance saturates to a finite value.
    By combining high-precision exact diagonalization calculations and analytical arguments, we observe that the imbalance does not scale when the occupied bands can be deformed to their Hermitian limit.
    This suggests a direct connection between the corresponding bulk topological invariants and the skin effect in many-body systems.
    Importantly, this relation also holds for interacting systems. 
\end{abstract}

\maketitle
Over the years, the interplay between non-Hermiticity and topology has gathered an immense interest, both from experimental and theoretical perspectives. 
Non-Hermitian (nH) Hamiltonians can serve as an effective description of open systems, accurately modeling various non-conservative classical and quantum platforms, including metamaterials~\cite{Barton2018, Brandenbourger2019, Ghatak2020, Scheibner2020, Zhou2020, Rao2021, Chen2021a}, optics and photonics~\cite{Regensburger2012, Peng2014, Zeuner2015, Feng2017, Zhou2018, Harari2018, Bandres2018, Pan2018, ElGanainy2018, Miri2019}, or electric circuits~\cite{Ezawa2019,Ezawa2019a, Ezawa2019b, Hofmann2020, Helbig2020}.
Within the framework of topological band theory, non-Hermiticity gives rise to exotic phenomena without Hermitian counterparts.
This includes exceptional points~\cite{Heiss2004, Leykam2017, MartinezAlvarez2018}  -- degeneracies where eigenstates coalesce --  and novel topological phases characterized by a winding of the eigenspectrum in the complex plane~\cite{Rudner2009, Esaki2011, Lee2016, Shen2018,Liu2019, Zhou2019, Kawabata2019, Zhang2020a}.

Another unique feature of nH systems is the anomalous localization of all eigenstates dubbed the non-Hermitian skin effect~\cite{Lee2016, Yao2018, Lee2019a, Lee2019, Borgnia2020}.
All single-particle eigenstates become exponentially localized at the boundaries of the system.
Generally, skin modes arise due to the breakdown of reciprocity, i.e., introducing asymmetric hoppings.
The conventional bulk-boundary correspondence is then broken, obscuring the prediction of universal boundary phenomena from the periodic bulk properties~\cite{Xiong2018, Kawabata2018, Gong2018, Yao2018a,Kunst2018, Herviou2019a, Jin2019, Song2019, Yokomizo2019, OzcakmakliTurker2019, Edvardsson2019, Zirnstein2021}.
Indeed, the phase diagram of the single particle Hamiltonian with periodic boundary conditions (PBC) can significantly differ from the one with open boundary conditions (OBC).
The skin effect is directly linked to the structure of the single-particle spectrum with PBC~\cite{Okuma2020, Zhang2020a} .
A non-zero winding of the PBC energy bands around any point in the complex plane guarantees that the related OBC modes are exponentially localized.
The corresponding OBC eigenenergies are (nearly all) located within these non-zero winding regions.
The existence of the skin effect was confirmed experimentally~\cite{Brandenbourger2019, Song2020, Scheibner2020, Helbig2020, Xiao2020, zhang2021acoustic, Liu2021}.
Recently, the notion of skin effect has been expanded by introducing different types of skin states~\cite{Yi2020} such as symmetry-enforced variants~\cite{Rui2019, Okuma2020, Yoshida2020}, reciprocal skin effect~\cite{Hofmann2020} in two or higher dimensions, higher-order skin effect~\cite{Kawabata2020, Okugawa2020, Fu2021, Palacios2021, Zhang2021, Zou2021}, or the critical skin effect~\cite{Li2020, Yokomizo2021}.

So far, studies of the skin effect have largely focused on the single-particle Hamiltonian.
Most of the experiments described by these effective nH Hamiltonians are either weakly interacting or non-interacting bosonic systems (e.g. quantum optics) or classical systems (e.g. the various types of circuits and active media).
Moreover, in Hermitian non-interacting systems, the properties of the many-body states can be straightforwardly deduced from those of the single-particle Hamiltonian; this is not the case for nH models.
Due to the non-orthogonality of the single-particle eigenmodes, the Pauli principle no longer acts trivially.
The fermionic repulsion reshapes the occupied orbitals so that no more than one fermion (equivalently one hard-core boson) populates each physical site.
Hence, the exponential localization of all fermions at a boundary becomes impossible~\cite{Lee2020, Liu2020}.
Recently, several authors have investigated the fate of the skin modes in the many-body context~\cite{Mu2020, Lee2020, Shen2021, Zhang2021a, Yoshida2021, Xi2021a}.
For instance, for interacting nH systems such as topological Mott insulators, the skin effect was observed in fermionic systems~\cite{Liu2020, Cao2021}, but not in the bosonic case~\cite{Zhang2020, Xu2020}.
Nonetheless, no general criterion for the survival of the skin effect has been derived so far.

In this Letter, we investigate the connection between topological properties of non-interacting nH systems and the skin effect in a many-body wave-function. 
The exponential localization of the skin modes naturally translates to an asymmetry of the many-body density profile that grows linearly with the system size.
Through numerical simulations, using high-precision exact diagonalization (ED) and density matrix renormalization group (DMRG), we show that the many-body eigenstates of OBC systems exhibit a transition from a regime with an infinite asymmetry in the thermodynamic limit to one where it saturates.
The critical anisotropy corresponds to a change in the topology of the \emph{periodic} single-particle spectrum, even though the open system remains gapped at all times.
In addition, we show that both scaling regimes survive disorder and interactions.

\begin{figure}
    \centering
    \includegraphics[width=0.95\columnwidth]{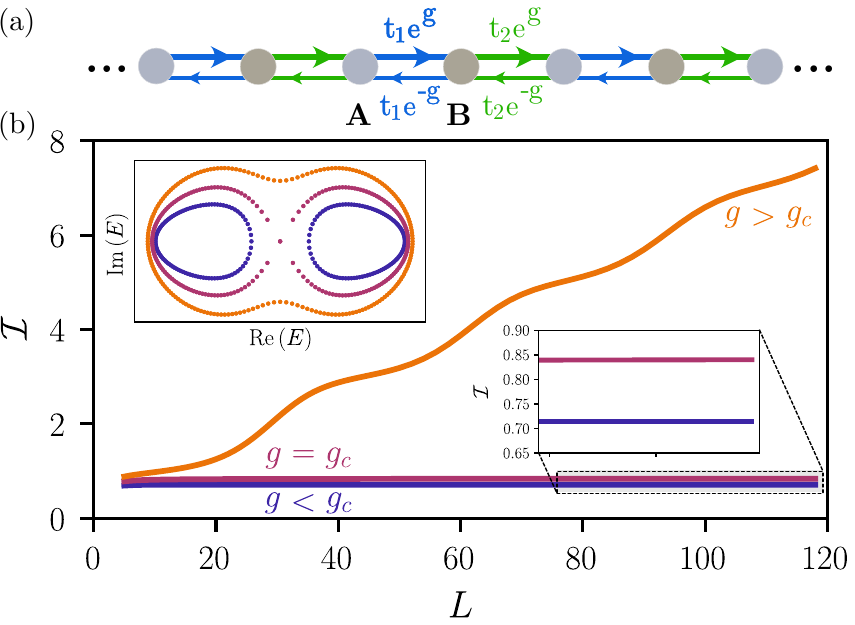}
    \caption{(a) Pictorial representation of the non-Hermitian SSH model. (b) Imbalance $\mathcal{I}$ as a function of the system size for $t_1/t_2 = 2$ and three representative values of $g$. At $g < g_c$, $\mathcal{I}$ saturates to a finite value. At $g > g_c$, the scaling becomes linear. It also exhibits a series of damped oscillations whose frequency diverges when $g \rightarrow g_c^+$. The inset shows the corresponding PBC spectra in the complex plane.}
    \label{fig:imbalance_noninteracting}
\end{figure}

\PRLSection{Conventions and model}
To exemplify our findings, we consider the nH chiral SSH model~\cite{Lieu2018, Kunst2018}:
\begin{multline}
    \mathcal{H}_\mathrm{SSH}\{g\} = -t_1 \sum\limits_{j} \left(e^{g} c^\dagger_{j, B} c_{j, A} + e^{-g} c^\dagger_{j, A} c_{j, B}\right) \\
    -t_2 \sum\limits_{j} \left(e^{g} c^\dagger_{j+1, A} c_{j, B} + e^{-g} c^\dagger_{j, B} c_{j+1, A}\right), \label{eq:SSHHamiltonian}
\end{multline}
where $c^{(\dagger)}$ are fermionic annihilation (creation) operators, $j$ denotes the lattice position, $A/B$ corresponds to the sublattice, and $t_{1/2}$ is the hopping amplitude.
$g\geq 0$ is a hopping anisotropy that breaks Hermiticity, see Fig.~\ref{fig:imbalance_noninteracting}(a).
We consider three boundary conditions: PBC, OBC (with $2L$ sites) or ABA-BC (OBC with $2L+1$ sites).
ABA-BC has an explicit analytical solution (see Appendix~\ref{app:analytical} and Ref.~\onlinecite{Kouachi2006}).
For PBC, the single-particle gap closes for $ \vert t_1 e^{\pm 2g} \vert = \vert t_2 \vert$, with two phases adiabatically connected to the trivial and topological Hermitian phases, and an intermediate phase where the system has no line gap.
In all cases, the single-particle spectrum has a non-trivial winding, and therefore the nH skin effect is always present.
Indeed, for OBC and ABA-BC, all single-particle eigenstates are exponentially localized to the right side of the system.
The model defined in Eq.~\eqref{eq:SSHHamiltonian} is related to the Hermitian SSH chain via the similarity transformation
\begin{equation}
    \mathcal{U}_g = \exp\left(g \sum\limits_j (2j-1) n_{j, A} + g\sum\limits_j 2j n_{j, B} \right)\label{eq:similaritytransform}
\end{equation}
such that
\begin{equation}
    \mathcal{H}_\mathrm{SSH}\{g\} = \mathcal{U}_g \, \mathcal{H}_\mathrm{SSH}\{0\} \, \mathcal{U}_g^{-1}.
\end{equation}
Note that the OBC and ABA-BC phase diagrams are therefore independent of $g$ with a gap closing at $t_1 = t_2$, and their spectrum is always real.

In the following, we focus on the ground state properties of the OBC and ABA-BC systems.
We define the ground state as the right eigenstate that minimizes the real part of the energy:
\begin{equation}
    \mathcal{H}_\mathrm{SSH}\{g\} \ket{\Psi\{g\}} = E_\mathrm{GS} \ket{\Psi\{g\}} \propto E_\mathrm{GS} \, \mathcal{U}_g\ket{\Psi\{0\}}.\label{eq:observableDefinition}
\end{equation}
Our results generalize straightforwardly to other right eigenstates and to the left eigenvectors.
The observables are taken to be:
\begin{equation}
    \braket{O}_g = \braket{\Psi\{g\}\vert O \vert \Psi\{g\}}.
\end{equation}
This expectation value is relevant when considering nH Hamiltonians obtained from post-selection~\cite{Dalibard1992, Moelmer1993}.
The quantum state described by such an approach remains a proper Hermitian density matrix.
As $\mathcal{H}_\mathrm{SSH}\{g\}$ is non-interacting, observables can be obtained efficiently from the diagonalization of the single-particle Hamiltonian~\cite{Herviou2019} (see Appendix~\ref{sec:decomposition}).
We numerically diagonalize the Hermitian system with high precision before applying the similarity transformation to obtain the eigenstates of the nH models.\\

\begin{figure}
    \centering
    \includegraphics[width=0.95\columnwidth]{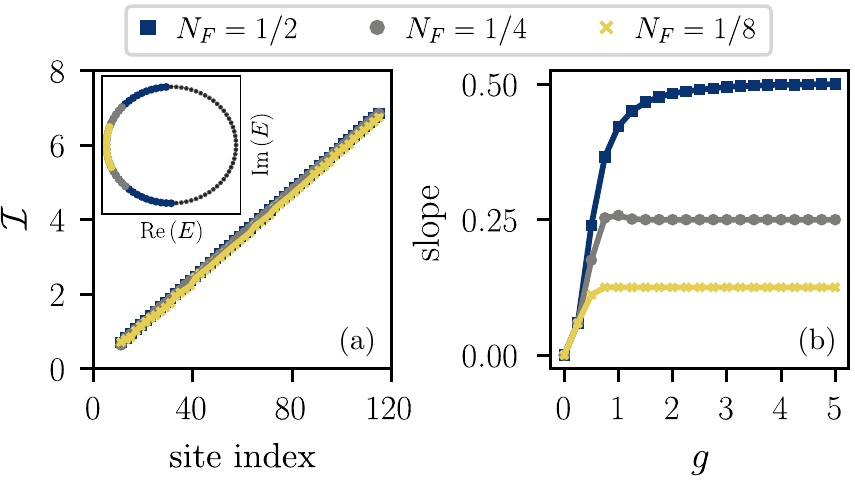}
    \caption{(a) Scaling of the imbalance in the Hatano-Nelson model with $t_1 = t_2 = 1$ and $g = 0.25$ at three different fillings $N_F$. Due to the structure of the PBC spectrum (see inset), the band is always partially occupied, which leads to a linear scaling with system size. (b) Slopes of the imbalance as a function of $g$. For small values of $g$, the slope does not depend on $N_F$. As $g$ is increased, all particles localize on the right boundary.}
    \label{fig:hatano_scaling}
\end{figure}

\PRLSection{Imbalance as a marker of the skin effect}
As can be seen from the transformation $\mathcal{U}_g$, all the single-particle orbitals are exponentially localized at the right boundary of the chain.
We expect the asymmetry of the orbitals to transfer to the many-body density distribution $\braket{n_{j}}_g = \braket{n_{j, A} + n_{j, B}}_g$.
Due to the Pauli principle, only one fermion can be located at a given site and therefore the many-body version of the skin effect is not a sum of the exponential orbitals.
To quantify the degree of asymmetry of the density distribution, we use the imbalance
\begin{equation}
    \mathcal{I} = \sum_{j \in \text{right half}} \braket{n_{j}}_g - \sum_{j \in \text{left half} } \braket{n_{j}}_g.
    \label{eq:imbalance}
\end{equation}
Working at fixed charge density $N_F = \sum\limits_j \braket{n_j}_g/2L \leq 0.5$\footnote{$N_F > 0.5$ can be deduced from $N_F < 0.5$}, the imbalance can at most scale linearly with system size.
We consider this linear scaling to be the manifestation of the skin effect in many-body systems.
In Fig.~\ref{fig:hatano_scaling}, we show the scaling of the imbalance in the limit $t_1 = t_2$ where the SSH model maps to the Hatano-Nelson model~\cite{Hatano1996, Mu2020}.
We observe a linear scaling of the imbalance at all fillings. 
At small $g$, the slope increases with the anisotropy $g$, but saturates at a value equal to $N_F$.
The large $g$ limit corresponds to a situation where all particles are located in the right half of the system.
We note that the imbalance can be generalized to the higher moments of the density with similar results.\\

\PRLSection{Skin effect and band topology}
We now turn to the general case where $t_1 \neq t_2$ and focus on half-filled systems ($L/(2L+1)$ filling for ABA-BC).
Fig.~\ref{fig:imbalance_noninteracting} summarizes our results for ABA-BC.
Below and at the critical value $g_c = 0.5 \, \vert\log t_2/t_1 \vert$, the imbalance saturates with system size.
For $g<g_c$, the saturating value is reached exponentially fast.
At the critical point, the convergence becomes algebraic, with an exponent which depends on the $t_1/t_2$ ratio and appears to peak at $t_1 = t_2$ (see Appendix \ref{sec:scaling} for more details).
Above $g_c$, the imbalance scales linearly with $L$, marking a fundamental anisotropy in the thermodynamic limit.
We also observe slowly decaying oscillations around the linear regime.
The relative amplitude of the oscillations, as well as their period, increases with decreasing $g$.
The limit $g \rightarrow g_c^+$ appears to correspond to a divergence of the period of the oscillation, as well as a vanishing of the linear term.
Due to the instability of the non-normal Hamiltonians, or alternatively the high amplitude of the coefficients in the similarity transformation $\mathcal{U}_g$, we stress the importance of using high-precision libraries to distinguish between finite-size~\cite{Chen2019} and finite-precision effects.
The system sizes shown here require a precision of up to several hundred of digits, hence the use of the ABA-BC where we can bypass diagonalization of the single-particle Hamiltonian.
The same results were obtained for OBC.\\

Remarkably, the critical value $g_c$ for the \emph{open} system corresponds to the gap closing of the \emph{periodic} system.
The single-particle Hamiltonian with OBC or ABA-BC remains gapped when crossing $g_c$\footnote{In fact the energy spectrum is independent of $g$}, and its eigenvectors vary smoothly. 
The abrupt change in the imbalance (discontinuous in the thermodynamic limit) directly arises from the Pauli principle.

To understand the source of the discontinuity, we summarize here how the many-body state is obtained from the single-particle orbitals.
Let us denote $\{ \ket{R_j} \}$ the right eigenvectors of the single-particle Hamiltonian, and $c^\dagger_{R, j} = \Vec{c}\,^\dagger \ket{R_j}$, the corresponding orbitals.
The eigenvectors of the many-body Hamiltonian are given by
\begin{equation}    
\ket{\Psi} \propto c^\dagger_{R, i_1} c^\dagger_{R, i_2}... \ket{0}.\label{eq:unormalizedstate}
\end{equation}
Due to the non-orthonormality of the $\{ \ket{R_{j}} \}$'s, the orbitals do not respect the conventional fermionic anticommutation relation and the right side of Eq.~\eqref{eq:unormalizedstate} is not normalized.
It is convenient to instead work with $\{ \ket{Q_{i_j}} \}$ --- an orthonormal family generating the occupied $\{ \ket{R_{i_j}} \}$ --- and the corresponding orbitals $\{ q^\dagger_{i_j} \}$.
They can be obtained by Gram-Schmidt orthonormalization.
We then have
\begin{equation}
    \ket{\Psi} = q^\dagger_{i_1} q^\dagger_{i_2}... \ket{0}.
\end{equation}
The $q_{i_j}$ now obey standard anticommutation relations.
Due to the orthonormalization, the smooth deformation of the original eigenvectors when $g$ crosses $g_c$ is amplified as the norm of the intermediate states can be arbitrarily close to zero.

We now focus on the saturating regime, $g < g_c$.
Note that the saturation of the imbalance depends on the fine structure of the orbitals.
Taking the corresponding PBC Hermitian state instead leads to a linear scaling of the imbalance after applying $\mathcal{U}_g$.
Generalized Brillouin zone (GBZ) approaches~\cite{Yao2018, Yokomizo2019, Yang2020, Kawabata2020a} give the same results.
It is informative to briefly study a simple example of saturation.
Consider $L$ independent orbitals 
\begin{equation}
    \ket{R_n} = \sum\limits_{j=1}^L a_{j, n}\ket{j}\otimes \ket{A} + b_{j, n} \ket{j}\otimes \ket{B}
    \label{eq:toymodel}
\end{equation}
such that  $a_{j, n} = \alpha b_{j, n}$ for $1 \leq n \leq L$.
By dimensional arguments, we obtain
\begin{equation}
    \begin{aligned}
\text{Span}(\ket{R_1}, ..., \ket{R_L}) = \text{Span}(\ket{1}\otimes (\ket{A} \\ + \alpha \ket{B}), ..., \ket{L}\otimes (\ket{A} + \alpha \ket{B})).
\label{eq:Space}
    \end{aligned}
\end{equation}  
The many-body state with $L$ occupied orbitals can therefore be rewritten as
\begin{equation}
   \ket{\Psi} = \frac{1}{(\sqrt{1+\alpha^2})^L} \prod_j (c^\dagger_{j, A} + \alpha c^\dagger_{j, B}) \Ket{0}
\end{equation}
independently of the exact form of $a_{j, n}$\footnote{We only require them to form an independent family.}.
This also holds when the orbitals are all exponentially localized at a boundary; the imbalance is constant and equal to zero.
As all orbitals have the same local structures, the Pauli principle forces electrons to spread out through the system.

The suppression of the scaling is observed only at half-filling in two-band models.
If the low-energy band is only partially occupied (resp. the high-energy band is partially occupied), the imbalance grows linearly with system size.
This can be immediately deduced from the toy model in Eq.~\eqref{eq:toymodel}: if we do not occupy the full translation-invariant subspace in Eq.~\eqref{eq:Space}, the imbalance survives.
In the nH-SSH model, for $g<g_c$, the periodic single-particle spectrum has a line gap, while for $g > g_c$, it forms a single band with no line gap and therefore its many-body spectrum is gapless.
Let $\{\mathcal{E}_j^\mathrm{PBC/OBC}\}$ be the set of energy bands with PBC or OBC, respectively, and the bands are separated by line gaps.
Following Ref.~\onlinecite{Okuma2020}, the OBC and PBC bands can be deformed into each other (merging several OBC bands to form a single one if required), up to some small set of states.
We propose the following conjecture:
\begin{conj}
For non-interacting systems exhibiting single-particle skin effect, if the sets of orbitals occupied in a many-body state $\ket{\Psi}$ can be mapped to fully occupied PBC bands (up to a number of orbitals of measure 0 in the thermodynamic limit), the many-body skin effect is suppressed.
Otherwise, the imbalance in $\ket{\Psi}$ will scale linearly with system size. 
\end{conj}

\PRLSection{Beyond the SSH model: disorder, interactions and symmetries}
Our conjecture implies that the many-body skin effect has a topological origin that can be predicted from considering the periodic system.
It is generally not the case for topological properties of nH systems which can be obtained from a GBZ approach or from the bulk of the OBC system.

\begin{figure}
    \centering
    \includegraphics[width=0.95\columnwidth]{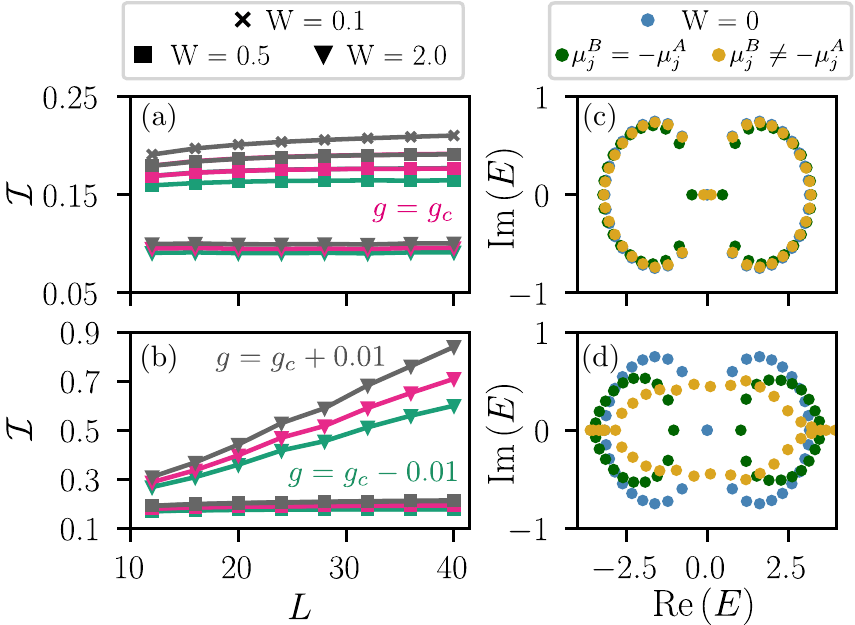}
    \caption{(a, b) Scaling of the imbalance for $t_1 / t_2 = 2$, $g = g_c$ and $g = g_c \pm 0.01$ for the two disorder configurations defined in Eq.~\eqref{eq:disorderdef} with W = $0.1, 0.5, 2$, averaged over $10 \, 000$ disorder realizations. (c, d) Typical PBC energy spectrum for the two disorder configurations at $g=g_c$ and (c) $W=0.5$ or (d) $W = 2$.
    The imbalance remains robust at small $W$ as the disorder opens a line gap. For (b), at $W = 2$, the line gap closes and the imbalance grows linearly.}
    \label{fig:disorder_imbalance_noninteracting}
\end{figure}

To further show the resilience of the skin effect, we investigate the effect of disorder~\cite{Hatano1996, Goldsheid1998, Yusipov2017, Tzortzakakis2020,Huang2020, Weidemann2021}.
For single-particle states, skin effect and disorder compete to localize states either at the boundaries or within the bulk of the system.
When the disorder-induced localization length is larger than the one induced by the breaking of reciprocity ($g^{-1}$ in our example), the single-particle states are Anderson-localized.
They then occupy arbitrary positions in the lattice which therefore limits the scaling of the imbalance.
On the other hand, the saturation is highly sensitive to the structure of the orbitals and therefore its survival is unclear.
In Fig.~\ref{fig:disorder_imbalance_noninteracting}, we show the imbalance in a disordered nH-SSH model with a random chemical potential
\begin{equation}
    -\sum\limits_j \mu_j^A  c^\dagger_{j, A}  c_{j, A} +  \mu_j^B  c^\dagger_{j, B}  c_{j, B} , \quad \mu_j^\alpha \in [ - W, W] \label{eq:disorderdef}
\end{equation}
for several values of the disorder strength $W$.
We consider two different disorder configurations: (a) $\mu_j^B = - \mu_j^A$, acting as a local chemical potential, or (b) $\mu_j^B$ and $\mu_j^A$ independent.
Saturated and linear regimes are resilient to the presence of small disorder in both setups.
For configuration (a), the disorder increases the critical value of $g$ by opening a line gap between the bands of the corresponding PBC models (see Fig.~\ref{fig:disorder_imbalance_noninteracting}(c, d)).
At small $W$, similar results are obtained for configuration (b).
However, at large $W$, the disorder closes the line gap at (and below) $g_c$ and the imbalance scales linearly accordingly.

\begin{figure}
    \centering
    \includegraphics[width=0.95\columnwidth]{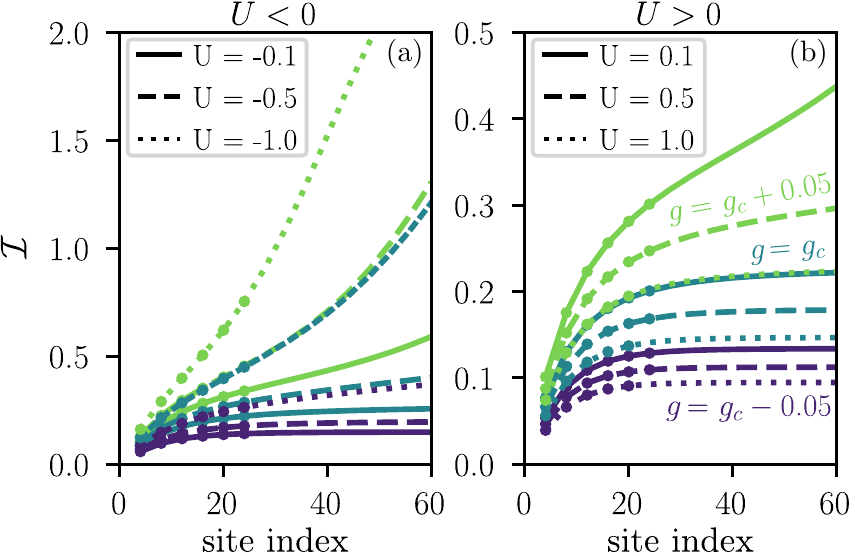}
    \caption{Imbalance scaling in the presence of (a) attractive and (b) repulsive interactions with $t_1 / t_2 = \sqrt{2}$. A smaller $t_1 / t_2$ ratio was used to reach larger sizes. The lines correspond to the DMRG data, while the superimposed points are the ED data up to 24 sites. Due to oscillations with a period four, we show the results only for every forth length. When $U <0$, the critical $g_c$ decreases and the slope of $\mathcal{I}$ increases. It is in contrast to positive $U$.}
    \label{fig:imbalance_interacting}
\end{figure}

In addition, we investigate the effect of interactions in the two scaling regimes.
We use DMRG~\cite{Tenpy} to access larger system sizes with OBC before applying the similarity transformation which is a matrix product operator of bond dimension $1$.
We are mainly limited by the floating point precision in the DMRG.
We add to the Hamiltonian in Eq.~\eqref{eq:SSHHamiltonian} the following interaction:
\begin{equation}
    \begin{aligned}
    H_\mathrm{int} = U\sum\limits_j \left( n_{j, A} - \frac{1}{2} \right) \left( n_{j, B} - \frac{1}{2} \right)& \\ +  \left( n_{j, B} - \frac{1}{2}\right) \left( n_{j+1, A} - \frac{1}{2} \right).
    \end{aligned}
    \label{eq:interactions}
\end{equation}
For $U$ real, the system remains gaugeable to a Hermitian model by the similarity transformation given in Eq.~\eqref{eq:similaritytransform}.
In the limit $t_1 = t_2$ (and $g=0$), the model is equivalent to the XXZ chain without transverse field.
In Fig.~\ref{fig:imbalance_interacting}, we show the scaling of the imbalance for (a) attractive and (b) repulsive interactions.
Both linear and saturated regimes survive the presence of small interactions.
For attractive interactions ($ U < 0$), fermions prefer to occupy neighbouring orbitals, and therefore the critical $g_c$ decreases.
Conversely, for repulsive interactions ($ U > 0 $), the interactions favor the electrons spreading throughout the system, so $g_c$ increases.
The notion of bands is no longer well-defined in the presence of interactions, but systems without a single-particle line gap have a gapless many-body spectrum. 
We verify that the variation of the critical $g_c$ matches a shift of the gap closing point.
We use the response of the periodic system to flux insertion to detect the presence of a gap given the limited system sizes we have access to.

In the Appendices, we show that our results are also valid when the system is no longer gaugeable to a Hermitian limit or when we break particle-hole symmetry by introducing a third band.\\

\PRLSection{Conclusions}
In this Letter, we have investigated the skin effect in many-body wavefunctions.
Generically, for fermionic systems, the single-particle skin effect translates into a linear scaling of the imbalance.
However, we have numerically shown that if the orbitals occupied in an OBC many-body state correspond to fully occupied PBC bands, the skin effect is suppressed in the thermodynamic limit.
Remarkably, this sharp transition occurs while the OBC spectrum remains completely gapped.
We have demonstrated that the suppression of the skin effect, despite being sensitive to the details of the orbitals, survives both the presence of disorder and interactions.
This strongly implies a topological origin of this phenomenon, similar to the topological nature of the nH skin effect in single-particle Hamiltonians.
An analytical proof of this connection to topology is left for future work.
Our results pave the way to a better understanding of the skin effect in higher dimensions, with the interplay between the dimensionality of the Fermi surface and other forms of skin effect.

\begin{acknowledgments}
The authors thank Jens Bardarson, Jérémy Bensadon, Tomáš Bzdušek, Frédéric Mila, Titus Neupert, Nicolas Regnault, and Songbo Zhang for the useful discussions.
Exact diagonalization studies were performed with QuSpin~\cite{Weinberg2017, Weinberg2019} and DMRG calculations were done using TeNPy Library (version 0.8.4)~\cite{Tenpy}. To achieve an arbitrary precision, we employed \texttt{mpmath} library~\cite{Johansson2021}. This work was supported by a grant from the Swiss National Supercomputing Centre (CSCS) under project s1008. F.A. acknowledges support from the KAUST Gifted Student Program and the EPFL Research Internship Program.
\end{acknowledgments}
\bibliography{skin_effect.bib}

\appendix
\section{Analytical solution for ABA-BC nH-SSH model}
\label{app:analytical}
The nH-SSH model with ABA-BC takes the form:
\begin{equation}
    H = \begin{pmatrix}
        0 & t_1 e^{-g} & 0 & 0 &\cdots & 0 \\
        t_1 e^g & 0 & t_2 e^{-g} & 0 & \cdots & 0 \\
        0 & t_2 e^g & 0 & \ddots & \ddots & \vdots \\
        0 & 0 & \ddots & \ddots & \ddots & 0 \\
        \vdots & \vdots & \ddots & \ddots & \ddots & t_2 e^{-g} \\
        0 & \cdots &\cdots & 0 & t_2 e^g & 0
    \end{pmatrix}.
    \label{eq:analyticalH}
\end{equation}
Following Ref.~\onlinecite{Kouachi2006}, we can write explicitly its eigenvalues and eigenstates.
Note that due to the choice of boundary conditions, there is always a zero-energy edge mode.
The eigenvalues are given by:
\begin{equation}
    E_n =  \begin{cases}
    \sqrt{t_1^2 + t_2^2 + 2 t_1 t_2 \cos \theta_n}, \,  &n = 1, \ldots, L \\
    - \sqrt{t_1^2 + t_2^2 + 2 t_1 t_2 \cos \theta_n}, \, &n = L + 1, \ldots, 2L \\
    0, \,   &n = 2 L + 1
  \end{cases} 
    \label{eq:analyticalE}
\end{equation}
and the corresponding eigenstates $u_j^{(n)}$ are
\begin{equation}
    u_j^{(n)} = \rho_j \begin{cases}
        t_1 t_2 \sin \left( \left[ \frac{2L + 1 -j}{2} + 1 \right] \theta_n \right) + t_1^2 \sin \left( \frac{2L + 1 - j}{2} \theta_n \right), \\
        \qquad \qquad \qquad \qquad \qquad \qquad \text{if } j \text{ is odd} \\
        -\sqrt{t_1 t_2} E_n \sin \left( \left[ \frac{2L + 1 -j}{2} + \frac{1}{2} \right] \theta_n \right), \\
        \qquad \qquad \qquad \qquad \qquad \qquad \text{if }  j \text{ is even}
    \end{cases}
    \label{eq:analyticalU}
\end{equation}
with 
\begin{equation}
    \rho_j = \begin{mycases}
        (\sqrt{t_1 t_2})^{2L} \, e^{g (j - 1)}, \, &j \text{ is odd} \\
        - (\sqrt{t_1 t_2})^{2L} \sqrt{\frac{t_1}{t_2}} \, e^{g (j -1)}, \, &j \text{ is even}
    \end{mycases}
    \label{eq:analyticalU2}
\end{equation}
and 
\begin{equation}
    \theta_n = \begin{cases}
        \frac{n \pi}{L + 1}, \, & 1 \leq n \leq L \\
        \frac{(n - L) \pi}{L+1}, \, & L +1 \leq n \leq 2L.
    \end{cases}
    \label{eq:analyticalTheta}
\end{equation}
The zero energy eigenstate is then given by
\begin{equation}
    u_{j}^{(2 L + 1)} = 
    \begin{mycases}
    \left(t_1 t_2 e^{2g} \right)^{(j-1)/2} \left( -t_2^2 \right)^{(2 L + 1 - j) / 2}, \, &j \text{ is odd} \\
    0, \, &j \text{ is even.} 
    \end{mycases}
    \label{eq:analyticalZeroMode}
\end{equation}

\section{QR decomposition for right eigenvectors}
\label{sec:decomposition}
We summarize here the main arguments of Ref.~\onlinecite{Herviou2019}.
We focus on the properties of the right eigenstates of the many-body Hamiltonian.
All observables are taken as in Eq.~\eqref{eq:observableDefinition}.
For non-interacting Hamiltonians, we can determine the properties of the ground states (and all eigenstates) by computing the correlation matrix~\cite{Peschel2003}.
The correlation matrix can be obtained from the eigenvectors of the single-particle Hamiltonian.
Let us consider a Hamiltonian of the form:
\begin{equation}
    \mathcal{H} = \vec{c}\,^\dagger H \vec{c}, 
\end{equation}
with $\vec{c}=(c_1, c_2, ...)$ the vector of annihilation operators whose index run over all degrees of freedom.
If $H$ is diagonalizable, with 
\begin{equation}
    H = \sum\limits_n E_n \ket{\psi_n^R} \bra{\psi_n^L},
\end{equation}
then we define the operators:
\begin{equation}
    \vec{r}_n^\dagger = \vec{c}\,^\dagger \ket{\psi^R_n}, \qquad \vec{l}_n^\dagger = \vec{c}\,^\dagger \ket{\psi^L_n}.
\end{equation}
They verify the biorthogonal fermionic anticommutation relations:
\begin{equation}
    \{r_m^\dagger, l_n \} = \delta_{m, n},
\end{equation}
\begin{equation}
    \{r_m, l_n \} = \{r_m, r_n \}=\{l_m, l_n \}=0.
\end{equation}
The many-body Hamiltonian $\mathcal{H}$ can be rewritten as 
\begin{equation}
    \mathcal{H} = \sum\limits_n E_n r_n^\dagger l_n.
\end{equation}
With this expression, we see directly that its right eigenstates are proportional to 
\begin{equation}
    \ket{\Psi} \propto \prod\limits_{j} r^\dagger_{i_j} \ket{0}.
\end{equation}
Due to the non-orthogonality of the diagonalizing basis of nH operators, these states are not normalized.
Moreover, contrary to the Hermitian case, observables cannot be directly computed from the expression above.
Instead, one needs to orthonormalize the family of occupied orbitals.
If $\{\ket{\phi_j}\}$ is an orthonormal basis of $\mathrm{Span}(\{ \ket{\psi^R_{i_j}} \}$, and we define 
\begin{equation}
    \vec{q}_n^\dagger = \vec{c}\,^\dagger \ket{\phi_n},
\end{equation}
 we see that the $q_{i_j}$'s are conventional fermionic operators.
The eigenstate in the form
\begin{equation}
    \ket{\Psi} = \prod\limits_{j} q^\dagger_{j} \ket{0} \propto \prod\limits_{j} r^\dagger_{i_j} \ket{0}
\end{equation}
is indeed normalized.
$\ket{\Psi}$ is a conventional fermionic Gaussian state, on which we can apply all the standard tricks.

\section{Scaling analysis}
\label{sec:scaling}
\begin{figure}[h]
    \centering
    \includegraphics[width=0.95\columnwidth]{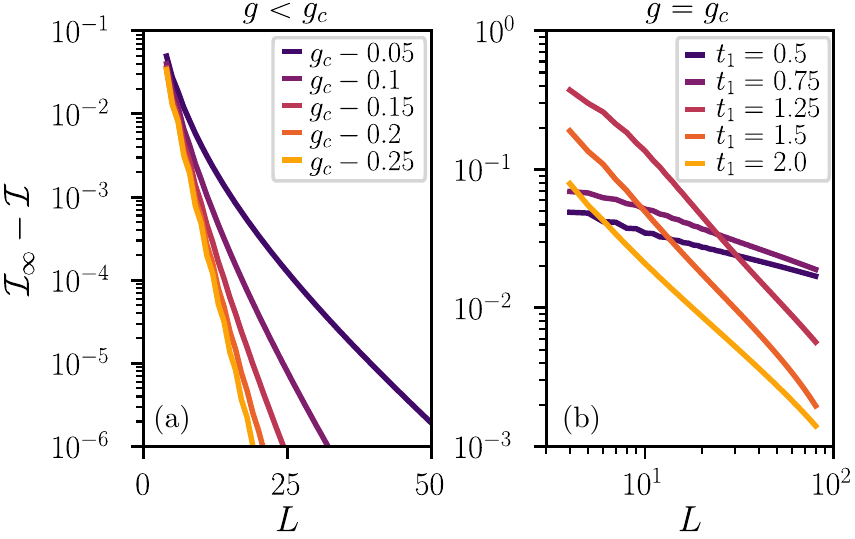}
    \caption{Scaling of the imbalance as a function of the system size $L$. (a) Below $g_c$, the imbalance saturates exponentially fast with a correlation length that varies with both $g$ and $t_1/t_2$. (b) At $g_c$, the convergence becomes algebraic with an exponent which depends on $t_1/t_2$.}
    \label{fig:scaling_saturation}
\end{figure}

As discussed in the main text, the imbalance as a function of the system size exhibits intriguing features depending on the value of $g$. 
To underline the universality of our results, we perform a scaling analysis for three $g$ regimes. 
Below $g_c$, the imbalance saturates to a finite and constant value. 
Away from the dimerized limits ($t_1= 0$ or $t_2 = 0$) and from the gapless point $t_1 = t_2$, the imbalance $\mathcal{I}$ can be well approximated by an exponential $\mathcal{I} (L) = \mathcal{I}_{\infty} + \lambda e^{-L/ \xi} $, as shown in Fig.~\ref{fig:scaling_saturation}(a).
At $g = g_c$, the correlations become algebraic with an exponent that depends on $t_1/t_2$, as presented in Fig.~\ref{fig:scaling_saturation}(b).

\begin{figure}[h]
    \centering
    \includegraphics[width=0.95\columnwidth]{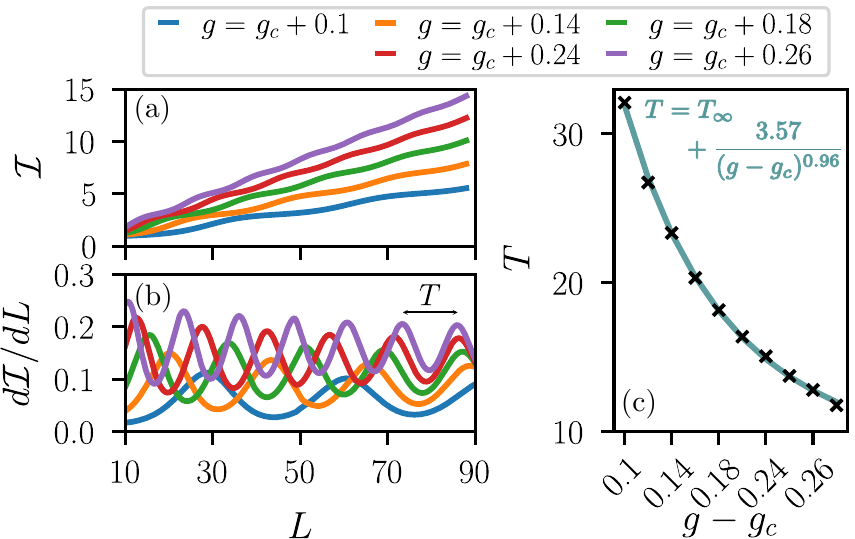}
    \caption{(a) Imbalance scaling for several $g >g_c$. Each curve exhibits a sequence of plateaus, with a period depending on the exact value of $g$. To quantify this dependence, we compute the numerical derivative $d \mathcal{I} / d L$ and perform a spline interpolation to estimate the positions of the peaks, shown in (b). The distance between peaks $T$  diverges as $\sim (g-g_c)^{-1}$.}
    \label{fig:scaling_oscillations}
\end{figure}

Finally, in the $g>g_c$ regime, we observe oscillations of the imbalance around the linear scaling (see Fig.~\ref{fig:scaling_oscillations}(a)).
To analyze these oscillations, we study the derivative $d \mathcal{I} / d L$, which clearly shows distinct peaks as demonstrated in Fig.~\ref{fig:scaling_oscillations}(b). 
The amplitude of the oscillations slowly decays with $L$, while their period appears to diverge as $(g -  g_c)^{-1}$: the saturation can be seen as an infinitely long plateau (see Fig.~\ref{fig:scaling_oscillations}(c)).

\section{Requirements for numerical precision}
\begin{figure}
    \centering
    \includegraphics[width=0.95\columnwidth]{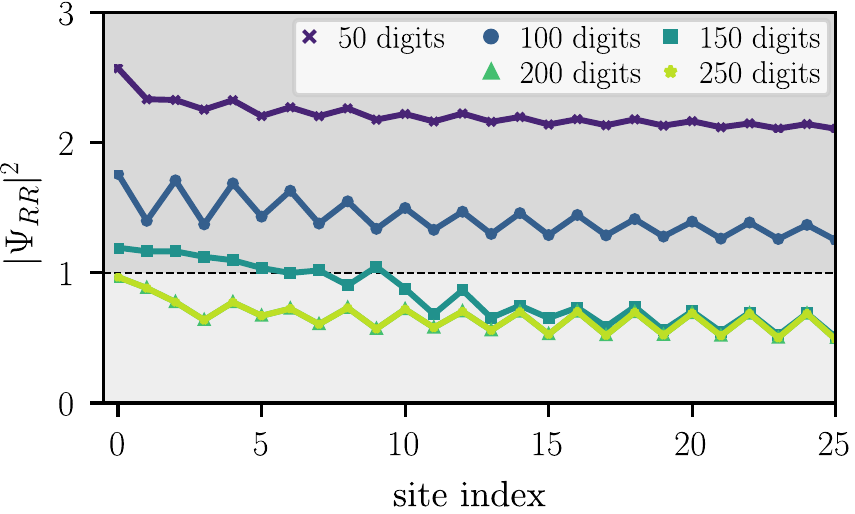}
    \caption{Many-body densities $|\Psi_{RR}|^2$ obtained from the analytical solutions for ABA-BC with $g = g_c + 0.1$ and $L = 400$ ($801$ sites) at different numerical accuracies in the Gram-Schmidt procedure. The numerical inaccuracies lead to incorrect density profiles at the edges of the system. For the given system size, at least 200 digits precision is required to faithfully represent the density. A standard double type variable uses 64 bits to store the value, which translates to around 16 decimal digits.}
    \label{fig:accuracy_comparison}
\end{figure}

Apart from the errors originating from too small system sizes, i.e., finite size effects,  numerical instabilities in nH matrices pose additional difficulties.
64-bits floating point precision becomes insufficient due to a combination of large system sizes and large values of the reciprocity-breaking terms. 
It is therefore necessary to systematically study how numerical accuracy affects the results, as it is known that nH matrices are exponentially unstable~\cite{Krause1994, Herviou2019a}.
In Fig.~\ref{fig:accuracy_comparison}, we compare the many-body ground state densities obtained from Gram-Schmidt orthonormalization at different numerical precisions. 
At low precision (up to $150$ digits), the local density close to the edge exceeds $1$, which is mathematically incorrect for normalized fermionic states.
The minimal number of required decimal digits scales roughly as $g L$.

\section{Non-gaugeable models}
So far, we showed that the relation between occupied orbitals and the scaling imbalance works for gaugeable models, i.e., models with real OBC spectrum on which we can apply a local similarity transformation.
The single-particle Hamiltonian (represented in momentum space)
\begin{equation}
    H(k) = H_\mathrm{SSH}(k) + \begin{pmatrix}
        0 & t_p e^{g_p} e^{2 \ii k} \\
        t_p e^{g_p} e^{-2 \ii k} & 0 \\
    \end{pmatrix}
    \label{eq:Hnongauge}
\end{equation}
is not gaugeable when $g_p \neq 3 g$. The bulk gap closes under the condition $t_p = (t_1 e^{\pm g} - t_2 e^{\mp g} ) e^{\mp g_p}$. 
We verified that the imbalance indeed changes scaling at the critical value above.

\section{Three-band models}
\begin{figure}
    \centering
    \includegraphics[width=0.95\columnwidth]{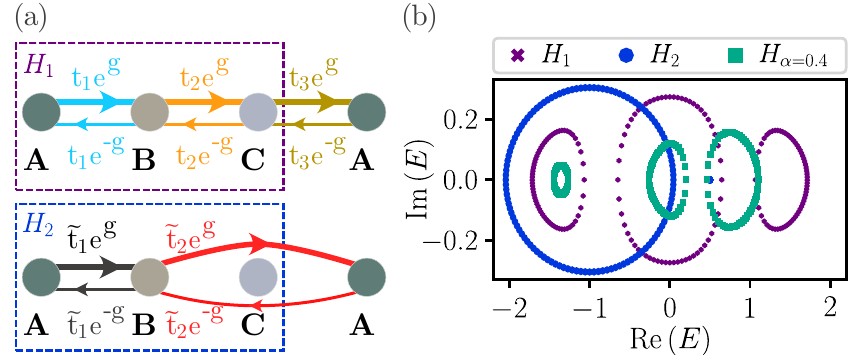}
    \caption{(a) A sketch of real-space hoppings for the two three-band models, $H_1$ and $H_2$, together with (b) their spectra. We set $t_1 = t_2 = 1$, $t_3 = 0.5$ for $H_1$ and $\tilde{t}_1 = \tilde{t}_2 = 0.5$, $\mu_1 = \mu_2 = -1$, $\mu_3 = 0.5$ for $H_2$; $g = 0.2 $. By linear interpolation between $H_1$ and $H_2$, we obtain a model for which the eigenvalues go from three circles to two.}
    \label{fig:three_bands}
\end{figure}

The existence of two distinct scaling regimes of $\mathcal{I}$ is not restricted to two-band models.
In this Appendix, we build a three-band model that can have either three bands separated by line gaps, or two asymmetric bands with a different number of orbitals. 
These regimes non-trivially break particle-hole symmetry.
To do so, we consider two elementary three-band models.
$H_1$ corresponds to a generalized nH-SSH model with three sites per unit cell:
\begin{equation}
    H_1 (k) = \begin{pmatrix}
        0 & t_1 e^g & t_3 e^{-g} e^{- \ii k} \\ 
        t_1 e^{-g} & 0 & t_2 e^g \\
        t_3 e^g e^{\ii k}  & t_2 e^{-g} & 0
        \end{pmatrix}.
    \label{eq:3bandsH1}    
\end{equation}
$H_2$ is a conventional nH-SSH model to which we attach an extra decoupled site within a unit cell:
\begin{equation}
    H_2 (k) = \begin{pmatrix}
        \mu_1 & \tilde{t}_1 e^g + \tilde{t}_2 e^{-2g} e^{-2 \ii k} & 0 \\
        \tilde{t}_2 e^{2g} e^{2 \ii k} + \tilde{t}_1 e^{-g} & \mu_2 & 0 \\
        0 & 0 & \mu_3
        \end{pmatrix}.
    \label{eq:3bandsH2}    
\end{equation}
We then study the interpolation $H = (1 - \alpha) H_1 + \alpha H_2$. 
The models and their eigenspectra are represented in Fig.~\ref{fig:three_bands}.
Whether we consider $1/3$, $1/2$ or $2/3$-filled ground states, the imbalance follows the conjecture in the main text for all parameter values.
In particular, we verified that the transition from saturated to linear regime occurs when the system goes from three separate bands (with PBC, $L$ orbitals per band) to two bands (one with $2L$ orbitals and another one with $L$ orbitals) in the $1/3$-filled case.

\section{Flux insertion in the nH periodic Hamiltonians}
\begin{figure}
    \centering
    \includegraphics[width=0.95\columnwidth]{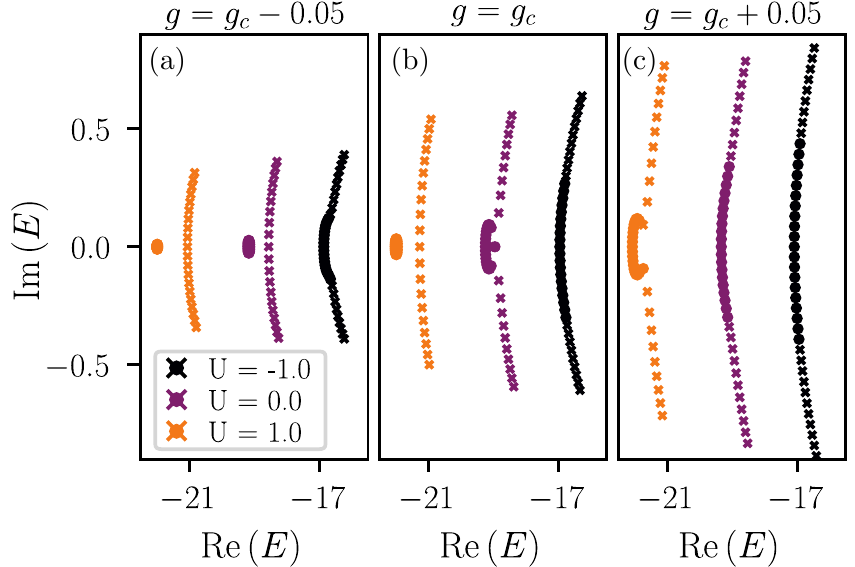}
    \caption{Evolution under flux insertion of the ground energy (dots) and first excited energy (crosses) in the model given in Eqs.~\eqref{eq:SSHHamiltonian} and \eqref{eq:interactions} with PBC  for (a) $g < g_c $, (b) $g = g_c$, and (c) $g > g_c $. The computations were performed for $U = -1$, $0$ and $1$ and at fixed system size $L = 12$ ($24$ orbitals).}
    \label{fig:manybody_spectra}
\end{figure}

To check whether our interacting periodic systems have a many-body gap, we use flux insertion.
Due to the limited system sizes and the lack of similarity transformation to map the systems back to Hermitian equivalents, scalings of the gap are unreliable.
Instead, we introduce a flux $\phi$ at the boundary of the system, and study the evolution of the lowest energies when we tune $\phi$ from $0$ to $2\pi$.
If the system is gapped, the lowest energy remains detached from the upper one (and its trajectory is a closed contour in the complex plane).
In the gapless case, the trajectory merges with those corresponding to higher energy and is no longer closed.
In Fig.~\ref{fig:manybody_spectra}, we show the trajectories of the two lowest energies in the model given in Eqs.~\eqref{eq:SSHHamiltonian} and \eqref{eq:interactions} with PBC for several values of $U$.
The change of behavior (from gapped to gapless, and vice-versa) matches the change in the scaling of the imbalance (from saturated to linear regime).

\section{Many-body computation of the imbalance}
In this Appendix, we briefly introduce some convenient formulas for computing the imbalance and its variants in many-body systems.
The imbalance $\mathcal{I}$ can be expressed as:
\begin{equation}
    \mathcal{I} = \frac{\braket{\Psi(0) \vert I \mathcal{U}_g^2\vert \Psi(0)}}{\braket{\Psi(0)\vert \mathcal{U}_g^2 \vert \Psi(0)}},
\end{equation}
where $\ket{\Psi(0)}$ is the Hermitian ground state.
From this structure, it is mathematically more convenient to work with
\begin{equation}
    \tilde{I}_1 = \frac{1}{L} \sum\limits_{j}\left( 2j-1 n_{j, A} + 2j n_{j, B} \right) - (2L-1).
\end{equation}
This is a minor generalization of the imbalance introduced in the main text, with essentially the same properties.
It allows us to write:
\begin{equation}
    \tilde{\mathcal{I}}_1 = \braket{\tilde{I}_1} = \frac{1}{2gL} \partial_g \log \braket{\Psi(0)\vert \exp(2g L \tilde{I}_1 ) \vert \Psi(0)}.
\end{equation}
$\tilde{\mathcal{I}}_1$ is therefore proportional to the derivative of the cumulant generating function of $L \tilde{I}_1$.
We numerically check that in the non-interacting case, the probability distribution is
\begin{equation}
    p(\tilde{I}_1 = i_1) \propto \exp\left(- \beta \vert L i_1 \vert \right) \exp\left(- \alpha i_1^2 \right)
\end{equation}
with 
\begin{equation}
    \beta = \vert \log t_1/t_2 \vert,
\end{equation}
which explains our results.


\end{document}